\documentclass[twocolumn]{aastex701}

\accepted{\today}

\begin{document}

\title{The JWST Proto-PAH Project: Detection of the methyl radical CH$_{3}$ in the highly evolved C-rich object SMP LMC 011}

\author[0000-0003-0665-6505]{Jialu Li}
\affiliation{Instituto de Astrof\'{\i}sica de Canarias, C/ Via L\'actea s/n, E-38205 La Laguna, Spain}
\affiliation{Departamento de Astrof\'{\i}sica, Universidad de La Laguna (ULL), E-38206 La Laguna, Spain}
\email{jialu.li@iac.es}

\author[0000-0002-1693-2721]{D. A. Garc\'{\i}a-Hern\'andez}
\affiliation{Instituto de Astrof\'{\i}sica de Canarias, C/ Via L\'actea s/n, E-38205 La Laguna, Spain}
\affiliation{Departamento de Astrof\'{\i}sica, Universidad de La Laguna (ULL), E-38206 La Laguna, Spain}
\email[show]{agarcia@iac.es}

\author[0000-0002-3011-686X]{A. Manchado}
\affiliation{Instituto de Astrof\'{\i}sica de Canarias, C/ Via L\'actea s/n, E-38205 La Laguna, Spain}
\affiliation{Departamento de Astrof\'{\i}sica, Universidad de La Laguna (ULL), E-38206 La Laguna, Spain}
\affiliation{Consejo Superior de Investigaciones Cient\'{\i}ficas (CSIC), Spain}
\email{amt@iac.es}

\author[0009-0001-4100-9218]{Debayan Das}
\affiliation{Department of Physics and Astronomy, The University of Western Ontario, London, ON N6A 3K7, Canada}
\email{ddas25@uwo.ca}

\author[0000-0002-2666-9234]{J. Cami}
\affiliation{Department of Physics and Astronomy, The University of Western Ontario, London, ON N6A 3K7, Canada}
\affiliation{Institute for Earth and Space Exploration, The University of Western Ontario, London, ON N6A 3K7, Canada}
\email{jcami@uwo.ca}

\author[0000-0002-2541-1602]{Els Peeters}
\affiliation{Department of Physics and Astronomy, The University of Western Ontario, London, ON N6A 3K7, Canada}
\affiliation{Institute for Earth and Space Exploration, The University of Western Ontario, London, ON N6A 3K7, Canada}
\email{epeeters@uwo.ca}

\author[0000-0003-4520-1044]{G. C. Sloan}
\affiliation{Space Telescope Science Institute, 3700 San Martin Drive, Baltimore, MD 21218, USA}
\affiliation{Department of Physics and Astronomy, University of North Carolina, Chapel Hill, NC 27599-3255, USA}
\email{gcsloan@stsci.edu}

\author[0000-0001-9848-5410]{B. Aringer}
\affiliation{Theoretical Astrophysics, Department of Physics and Astronomy, Uppsala University, Box 516, 751 20 Uppsala, Sweden}
\affiliation{Department of Astrophysics, University of Vienna, Türkenschanzstraße 17, 1180 Wien, Austria}
\email{bernhard.aringer@aon.at}

\author[0000-0002-8452-8675]{J. Bernard-Salas}
\affiliation{ACRI-ST, Centre d’Etudes et de Recherche de Grasse (CERGA), 10Av. Nicolas Copernic, 06130 Grasse, France}
\affiliation{INCLASS Common Laboratory, 10 Av. Nicolas Copernic, 06130 Grasse, France}
\email{jeronimo.bernard-salas@acri-st.fr}

\author[0009-0000-0191-6756]{C. Bhatt}
\affiliation{Department of Physics and Astronomy, The University of Western Ontario, London, ON N6A 3K7, Canada}
\affiliation{Institute for Earth and Space Exploration, The University of Western Ontario, London, ON N6A 3K7, Canada}
\email{cbhatt7@uwo.ca}

\author[0009-0009-4643-2734]{Nicholas Clark}
\affiliation{Department of Physics and Astronomy, The University of Western Ontario, London, ON N6A 3K7, Canada}
\affiliation{Institute for Earth and Space Exploration, The University of Western Ontario, London, ON N6A 3K7, Canada}
\email{nclark68@uwo.ca}

\author[0000-0002-4017-5572]{Harriet L. Dinerstein}
\affiliation{Department of Astronomy, University of Texas at Austin, 2515 Speedway, Stop C1400, Austin, TX, 78712-1205, USA}
\email{harriet@astro.as.utexas.edu}

\author[0000-0002-3938-4211]{M. A. G\'omez-Mu\~{n}oz}
\affiliation{Departament de Fïsica Quàntica i Astrofísica (FQA), Universitat de Barcelona (UB), C/ Martí i Franqués 1, E-08028 Barcelona, Spain}
\affiliation{Institut de Ciències del Cosmos (ICCUB), Universitat de Barcelona (UB), C/ Martí i Franqués 1, E-08028 Barcelona, Spain}
\email{mgomez@icc.ub.edu}

\author[0000-0002-2626-7155]{Kathleen E. Kraemer}
\affiliation{Institute for Scientific Research, Boston College, 140 Commonwealth Avenue, Chestnut Hill, MA 02467, USA}
\email{kathleen.kraemer@bc.edu}

\author[0000-0002-5529-5593]{M. Matsuura}
\affiliation{Cardiff Hub for Astrophysics Research and Technology (CHART), School of Physics and Astronomy, Cardiff University, The Parade, Cardiff CF24 3AA, UK}
\email{matsuuram@cardiff.ac.uk}

\author[0000-0002-6858-5063]{R. Sahai}
\affiliation{Jet Propulsion Laboratory, MS 183-900, 4800 Oak Grove Dr., California Institute of Technology, Pasadena, CA 91109, USA}
\email{raghvendra.sahai@jpl.nasa.gov}

\author[0000-0002-9604-1434]{N. C. Sterling}
\affiliation{University of West Georgia, 1601 Maple Street, Carrollton, GA
30118, USA}
\email{nsterlin@westga.edu}

\author[0000-0002-3824-8832]{Kevin Volk}
\affiliation{Space Telescope Science Institute, 3700 San Martin Drive, Baltimore, MD 21218, USA}
\email{volk@stsci.edu}

\author[0000-0002-6570-4776]{G. M. Wahlgren}
\affiliation{Space Telescope Science Institute, 3700 San Martin Drive, Baltimore, MD 21218, USA}
\email{gwahlgren@stsci.edu}

\author[0000-0002-3171-5469]{A. A. Zijlstra}
\affiliation{Jodrell Bank Centre for Astrophysics, Department of Physics \& Astronomy, The University of Manchester, Oxford Road, Manchester M13 9PL, UK}
\email{albert.zijlstra@manchester.ac.uk}


\begin{abstract}
We report the first detection of the neutral methyl radical CH$_3$ in {a} highly evolved C-rich object, SMP\,LMC\,011, using JWST MIRI/MRS. CH$_3$ is well fitted by an excitation temperature of $T_{\rm ex} \simeq$~190\,K and a column density of $N_{\rm tot}(\mathrm{CH_3}) \simeq 5.6\times10^{17}$ cm$^{-2}$. We also report the non-detection of ethane (C$_2$H$_6$), {consistent with} CH$_3$ {reacting} preferentially with unsaturated radicals rather than {self-recombination, which may be inefficient in SMP\,LMC\,011.} Combined with the exceptionally large benzene column density of SMP\,LMC\,011, this supports a methyl-addition route from benzene towards alkyl-substituted aromatics. We propose that the high CH$_3$ abundance is driven by the erosion of hydrogenated amorphous carbon (HAC) dust grains in the dense warm torus by UV photons from the central star and/or by shocks, which release CH$_3$ directly into the gas phase. These results establish CH$_3$ as a key reactive intermediate in the formation of complex hydrocarbons in C-rich circumstellar environments. Chemical models {that incorporate methyl-addition reactions are needed} for a better understanding of PAH formation pathways in evolved stars. The potential detection of alkyl-substituted aromatics such as toluene (C$_7$H$_8$) and ethylbenzene (C$_8$H$_{10}$) in C-rich evolved stars would provide direct confirmation of CH$_3$-driven aromatic growth in circumstellar environments.
\end{abstract}

\keywords{\uat{Small molecules}{2267} --- \uat{Circumstellar gas}{238} --- \uat{Astrochemistry}{75} --- \uat{Interstellar medium}{847} --- \uat{Planetary nebulae}{1249}}


\section{Introduction} 

Carbon-rich pre-planetary nebulae (PPNe) and young PNe are extraordinary chemical laboratories. As C-rich stars evolve from the asymptotic giant branch (AGB) toward the PN phase, the increasing UV radiation field from the warming central star transforms the circumstellar chemistry. Photochemical reactions in the neutral layers of the photodissociation region (PDR) are expected to produce high abundances of polyynes (C$_{2n}$H$_2$), cyanopolyynes (HC$_{2n+1}$N), methylpolyynes (CH$_3$C$_{2n}$H), and crucially benzene (C$_6$H$_6$), the basic building block of polycyclic aromatic hydrocarbons \citep[PAHs; e.g.,][]{cernicharo2001a,cernicharo2001b,cernicharo2004}. Revealing the reactive intermediates that drive the molecular chemistry is central to understanding the formation and growth of the complex C-rich molecules (e.g., PAHs) observed from Galactic objects to very distant galaxies.

A key species in this chemistry is the methyl radical, CH$_{3}$. As a highly reactive open-shell intermediate in interstellar ion-molecule reaction networks, CH$_{3}$ is believed to be formed via hydrogen abstraction reactions starting from C$^+$ and H$_2$ \citep[e.g.,][]{herbst1973} and through photodissociation of methane \citep[e.g.,][]{feuchtgruber2000}\footnote{{The ion-molecule initiation is endothermic and requires high temperatures or vibrationally excited H$_2$ to proceed; in the UV-irradiated torus of SMP LMC 011, gas-phase CH$_3$ is therefore expected to be produced mainly by methane photodissociation (see text).}}. Thus, it may play a dual role. Its reactions with C$^+$ represent one of the most important steps in the build-up of complex hydrocarbons, while its addition to aromatic radicals such as \textit{ortho}-benzyne (o-C$_6$H$_4$) provides a bottom-up pathway to five-membered rings \citep{bouwman2023} and alkyl-substituted aromatics \citep{Shukla2010,santoro2020}. The CH$_{3}$ radical is also expected to play a key role in HCN formation in PDRs \citep[e.g.,][]{Sternberg1995}. 

Despite this central role, CH$_{3}$ is notoriously difficult to detect. Being planar ($D_{3h}$ symmetry), it has no permanent electric dipole moment and no accessible pure rotational spectrum. Its detection has mainly relied on the $\nu_2$ out-of-plane bending mode near 16.5\,$\mu$m, a spectral window blocked by the Earth's atmosphere. The first detection of CH$_{3}$ was made with the Infrared Space Observatory (ISO) toward Sagittarius~A* in the Galactic center \citep{feuchtgruber2000}. In the JWST era, CH$_{3}$ has been identified in the protoplanetary disk around the very low-mass star ISO-ChaI 147 \citep{Arabhavi2024} and the extragalactic buried nucleus of IRAS~07251$-$0248 \citep{garciabernete2026}. Beyond the neutral radical, the methyl cation CH$_3^+$ has {now} been detected in {several} UV-irradiated environments, including a protoplanetary disk in the Orion star-forming region \citep{berne2023,changala2023}, {the Orion Bar PDR \citep{zannese2026}, the transition disks TW~Hya \citep{henning2024}, and GM~Aur \citep{volz2026} as well as} the O-rich PN NGC~6302 \citep{bhatt2025}. However, a detection of the neutral methyl radical CH$_{3}$ in a C-rich evolved stellar environment has remained elusive until now.
 
SMP\,LMC\,011 is uniquely suited for this search. It has been classified as a young PN in the Large Magellanic Cloud \citep[LMC;][]{sanduleak1978,bernardsalas2006}. Its Spitzer Infrared Spectrograph \citep[IRS;][]{Houck2004} spectrum reveals an exceptionally rich molecular absorption spectrum, including C$_2$H$_2$, C$_4$H$_2$, CH$_4$, C$_2$H$_4$, HC$_3$N, propyne (CH$_3$C$_2$H), and most remarkably benzene (C$_6$H$_6$; \citealt{bernardsalas2006, malek2012}). SMP\,LMC\,011 is one of only two highly evolved C-rich objects in which benzene has been detected (the other being the Galactic PPN AFGL\,618, the Westbrook Nebula; \citealt{cernicharo2001a}), and it harbors by far the largest known benzene column density \citep[$N_{\rm tot}(\mathrm{C_6H_6}) \simeq 6.3\times10^{17}$ cm$^{-2}$;][]{malek2012}, with a benzene-to-acetylene ratio roughly 20 times higher than in AFGL\,618 \citep{malek2012}. 

These anomalous abundances, inconsistent with current photochemical models of C-rich PPN outflows, point to unusual chemistry operating within a dense, warm, edge-on torus \citep{cernicharo2004, malek2012}. Hubble Space Telescope imaging by \citet{Shaw2006} revealed a compact bipolar morphology analogous to AFGL\,618, and the central star \citep[or possibly a binary system; see][hereafter Paper~\textsc{i}]{sloan2026} is entirely obscured by the optically thick dusty torus, with the total mid-IR flux dominated by warm dust emission \citep{malek2012}. 

The presence of methyl-bearing species such as CH$_3$C$_2$H, combined with the high column density of benzene, makes SMP\,LMC\,011 the most compelling environment in which to search for CH$_{3}$ as the key reactive intermediate linking small hydrocarbons to the aromatic chemistry. Here we report the first detection of the methyl radical CH$_{3}$ in SMP\,LMC\,011 using JWST MIRI/MRS observations, which represents the first {secure detection and quantitative characterization} of this species in an evolved stellar object. 




\section{JWST Observations and Data Reduction} \label{sec:jwst_data}
 
SMP\,LMC\,011 was observed on 2025 May 27 under JWST GO Cycle~3 program 4678, using the Medium-Resolution Spectrometer \citep[MRS;][]{argyriou2023} of the Mid-Infrared Instrument \citep[MIRI;][]{wright2023}. All three grating settings were employed to achieve continuous spectral coverage from 4.9 to 28.7 $\mu$m, and a 4-point dither pattern optimized for point sources was used. The program was designed to reach a signal-to-noise ratio of 100 or better over the 4.9--20 $\mu$m range. Full details of the observing strategy are given in Paper~\textsc{i}.

SMP\,LMC\,011 is unresolved in the IFU data, and the spectra were reduced using a development version of the JWST calibration pipeline (v1.20.2dev) together with the point fixed-pattern correction \citep[PFPC;][]{gordon2026}. The PFPC extracts spectra from each of the four dither positions independently and calibrates them against JWST standard star observations\footnote{All SMP\,LMC\,011 spectra presented here were calibrated using standard stars from the CALSPEC database \citep{gordon2022}.}. This approach effectively removes residual fringing and flat-field errors from the extracted spectra (see Paper~\textsc{i} for details). Discontinuities between the twelve spectral sub-bands were corrected by applying scalar multiplicative factors in the overlap regions between adjacent segments, propagating outward in both directions from Band~3A (12.57 $\mu$m); these stitching corrections were generally smaller than 1\%. 

We constructed a continuum-normalized MIRI/MRS spectrum over 5--20~$\mu$m for the molecular analysis presented below. The continuum was estimated with a smooth spline interpolation through selected continuum windows outside the strongest molecular absorption bands. Specifically, the broad absorption complexes at 6.65--8.35~$\mu$m and 12.0--17.5~$\mu$m were excluded from the continuum fit, together with the narrower absorption region at 10.25--10.85~$\mu$m. The observed spectrum was then divided by the fitted continuum, and the resulting normalized flux and propagated uncertainties were used for the molecular identification and spectral modeling described in Section~\ref{sec:analysis}.

\section{Molecular Identification and Spectral Modeling}\label{sec:analysis}

\subsection{Model Construction}

We adopted the spectroscopic parameters for CH$_3$ and C$_2$H$_6$ from the HITRAN database \citep{Rothman2021} and modeled their molecular absorption bands using the foreground-slab model, which is commonly used for MIR absorption spectra. We neglect the source-function contribution from the slab itself and assess the impact of this approximation in Section~\ref{subsec:results}. For an isothermal slab covering a fraction $f_c$ of the background continuum, the observed intensity is

\begin{equation}
I_\nu = I_{c}\left[1 - f_c (1-e^{-\tau_\nu})\right], \label{eq:1}
\end{equation}
where $I_{c}$ is the continuum intensity and $\tau_\nu$ is the optical depth of the absorbing gas. For a given excitation temperature $T_{\rm ex}$, total column density $N_{\rm tot}$, Doppler parameter $b$, and velocity shift~\citep[263.5~km~s$^{-1}$ for the LMC;][]{MP1998}, we computed the lower-state column density, $N_l$, of each transition, assuming LTE:
\begin{equation}
N_{l} =
N_{\rm tot} \frac{g_{l}}{Q(T_{\rm ex})}
\exp\left(-\frac{E_{l}}{k_B T_{\rm ex}}\right),
\end{equation}
where $g_{l}$ and $E_{l}$ are the lower-state statistical weight and energy, $Q(T_{\rm ex})$ is the partition function, and $k_B$ is the Boltzmann constant. 

The optical-depth profile of each transition was calculated from \begin{equation}
\tau_{\nu} = \tau_{p}\phi_{\nu} =
\sqrt{\pi} e^2/(m_e c)
 N_{l} f_{lu} (\lambda/{b})\phi_{\nu},
\end{equation}
where $e$ is the electron charge, $m_e$ is the electron mass, $c$ is the speed of light, $f_{lu}$ is the oscillator strength, and $b$ is the Doppler parameter in velocity space~\citep{Tielens2021}. $\tau_{p}$ is the peak optical depth and the profile function $\phi_{\nu}$ is a Voigt profile normalized to unity at line center, $\phi_{\nu_0}$=1, and consists of a Doppler core and a negligible Lorentzian damping wing~\citep{li2024}. The model spectrum was then calculated using Equation~\ref{eq:1} and convolved to {the wavelength-dependent MIRI/MRS resolving power. We adopted the linear resolving-power relations $R(\lambda)=A+B\lambda$ given by~\citet{pontoppidan2024}(their Table 3) using the appropriate coefficients for Channels 3C and 4A over the corresponding wavelength intervals. The convolution was performed locally with a Gaussian line-spread function whose width varies with wavelength according to the adopted $R(\lambda)$.}

\begin{figure*}
    \centering
    \includegraphics[width=\textwidth]{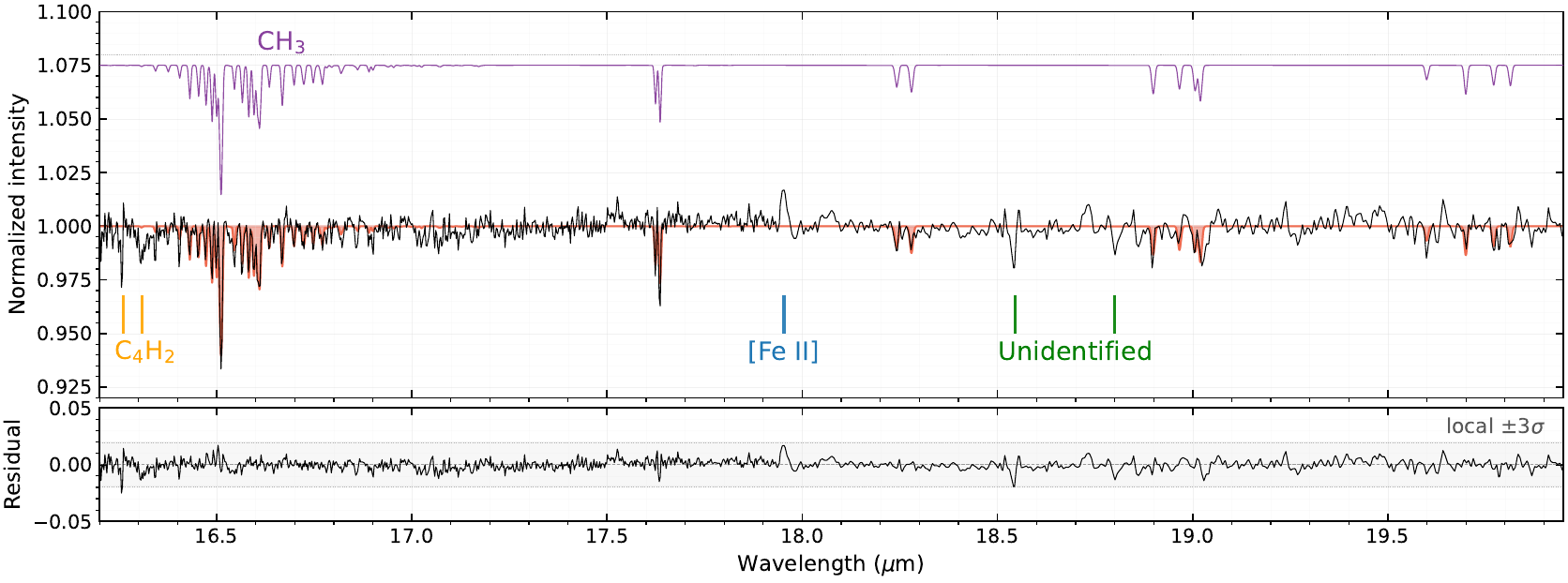}
    \caption{
    \textit{Top panel}: Continuum-normalized 16.2--20~$\mu$m JWST/MIRI MRS spectrum of SMP~LMC~011 (black), compared with the adopted CH$_3$ model (red shaded region). The offset purple spectrum shows the CH$_3$ model separately. The orange markers indicate the C$_4$H$_2$ absorption after subtracting the broad C$_4$H$_2$ plateau (see Figure~\ref{fig:ch3_c6h2_model_plateau_removel} in the Appendix), {the blue marker marks [Fe II] identified in} Paper~\textsc{i}, and the green markers denote unidentified absorption features.  \textit{Bottom panel}: residual after subtracting the CH$_3$ model from the normalized spectrum. The gray band indicates the local $\pm 3\sigma$ uncertainty, estimated as three times the median normalized MRS uncertainty over the 17--20~$\mu$m region.}
    \label{fig:ch3_c6h2_model_v2}
\end{figure*}

\subsection{Line Identification and Physical Constraints}\label{subsec:results}

The 16.2--20~$\mu$m region provides the clearest window for identifying CH$_3$ in the MIRI/MRS spectrum, although CH$_3$ has predicted absorption features throughout the broader 12--20~$\mu$m range. Compared with the more complex 12--16~$\mu$m region (Paper~\textsc{i}; Das et al., in prep.), the 16.2--20~$\mu$m spectrum contains several groups of recognizable CH$_3$ absorption features extending from $\sim$16.3 to 20~$\mu$m, together with a broad C$_4$H$_2$ plateau below 17~$\mu$m {\citep[previously identified in SMP~LMC~011,][]{bernardsalas2006,malek2012},} and several absorption features still unidentified here. Figure~\ref{fig:ch3_c6h2_model_v2} compares our representative CH$_3$ model with the continuum-normalized MRS spectrum.

\textbf{CH$_3$:} The line identification and model comparison were carried out iteratively. We first used the predicted CH$_3$ line positions and relative strengths to identify wavelength intervals in which CH$_3$ is expected to contribute. The adopted intervals are 16.3--16.9, 17.6--17.65, 18.2--18.32, 18.8--19.05, and 19.55--19.85~$\mu$m. These intervals avoid the short-wavelength region most strongly affected by the C$_4$H$_2$ plateau and exclude near-continuum regions that contain little molecular signal and are more sensitive to continuum uncertainties.

Within these intervals, we selected a fixed set of diagnostically important absorption features and compared the observed and modeled fluxes in narrow windows centered on their local minima. We then performed a grid search over $T_{\rm ex}$=50--350~K, log$[N_{\rm tot}\rm{(CH_3)/\rm{cm}^{-2}}]$=16.0--19.5, $b$=5--15~km~s$^{-1}$, and $f_c$=0.1--0.25, using a feature-based $\chi^2$-like statistic that measures how well each model reproduces the relative depths of the selected CH$_3$ features. {Appendix~\ref{app:uncertainties} describes the empirical motivation for the adopted $b$ and $f_c$ ranges, together with the details of the statistic, uncertainty treatment, and parameter degeneracies.} Because the fit quality changes only modestly along these degeneracies, we adopt the global minimum of the feature-based statistic as a representative solution rather than as a unique physical model. 

{This solution has $T{\rm ex}=190^{+25}_{-20}$~K, $\log[N_{\rm tot}({\rm CH_3})/{\rm cm}^{-2}]=17.75^{+0.05}_{-0.05}$, $b=10$~km~s$^{-1}$, and $f_c=0.1$ (Appendix~\ref{app:uncertainties}).} The quoted uncertainties are the projections of the joint 68.3$\%$ $T_{\rm ex}$ and log$N_{\rm tot}$ region for this fixed ($b$, $f_c$) pair. {They therefore represent only the formal uncertainties within this adopted model and do not include the broader systematic uncertainty associated with the poorly constrained $b$, $f_c$, continuum placement, and model assumptions.} For the representative solution adopted here, the omitted source-function term changes the continuum-normalized intensity by at most $\sim$1\% in the optically thick limit, and by less for unsaturated lines. For this solution, the strongest individual CH$_3$ transition in the fitted wavelength range has a peak optical depth of $\tau_p \simeq 11.5$ at 16.513 $\mu$m. {We note that this fitted $T_{\rm ex}$ characterizes the rotational population sampled by the $\nu_2$ absorption band rather than providing a direct measurement of the gas kinetic temperature. Both collisional redistribution and radiative pumping by the mid-IR continuum can contribute to setting the CH$_3$ rotational populations, so $T_{\rm ex}$ cannot be identified uniquely with $T_{\rm kin}$ without a full non-LTE treatment, which is beyond the scope of this paper.}

The present constraints on $b$ and $f_c$ are not independent of the adopted CH$_3$ model. A future simultaneous analysis of multiple molecular species or spectral bands in this data set may provide tighter and more physically motivated constraints on these parameters and reduce the associated degeneracies (e.g., Das et al., in prep.).

\textbf{C$_2$H$_6$:} C$_2$H$_6$ has several bands from the near- to the mid-IR. However, the C–H stretching bands near 3.3 $\mu$m fall within the NIRSpec spectral range analyzed by Das et al.\ (in prep.). On the other hand, the CH$_3$-deformation band near 6.8~$\mu$m is heavily blended with C$_6$H$_6$ and HCN absorption \citep[][Das et al., in prep.]{malek2012}. We therefore focus on the $\nu_9$ band near 12.2~$\mu$m, which provides the cleanest constraint on C$_2$H$_6$.

\begin{figure}[h]
    \centering
    \includegraphics[width=0.45\textwidth]{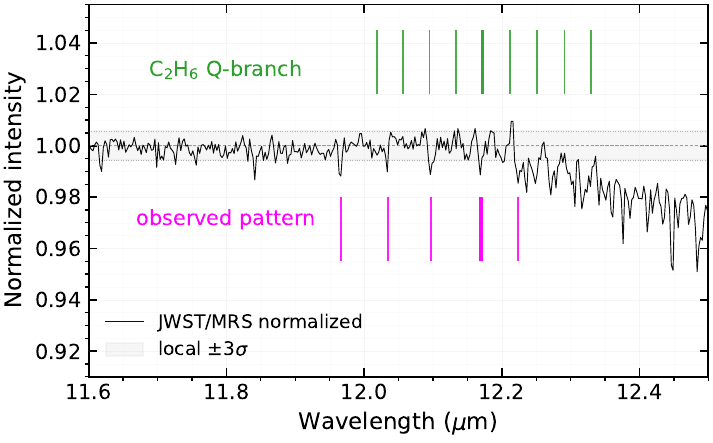}
    \caption{The 11.6--12.5~$\mu$m JWST/MIRI MRS spectrum of SMP LMC~011 (solid black), with the gray band indicating the local $\pm3\sigma$ uncertainty. Green vertical markers show the predicted positions of the C$_2$H$_6$ Q-branch absorption features, while magenta markers show the observed quasi-regular narrow absorption pattern. The thicker markers indicate the predicted central Q-branch minimum near 12.172~$\mu$m (green) and the nearest observed feature at 12.170~$\mu$m (magenta). Although this central observed feature lies close to the expected Q-branch center, the spacing of the observed pattern does not match the predicted C$_2$H$_6$ Q-branch structure, supporting the non-detection of C$_2$H$_6$.}
\label{fig:c2h6_12um}
\end{figure}

An absorption feature at {12.170} $\mu$m with a depth of $\sim$1.14\% is present near the predicted central Q branch of the C$_2$H$_6$ $\nu_9$ band at 12.172~$\mu$m, but the local pattern is inconsistent with C$_2$H$_6$ (Figure~\ref{fig:c2h6_12um}). {A search of the relevant spectroscopic databases and literature did not yield a convincing identification. If the quasi-regular spacing of $4.41\pm0.57~{\rm cm^{-1}}$ corresponds to consecutive P- or R-branch lines of a near-rigid linear molecule, it would imply a rotational constant of approximately $B\sim2.2~{\rm cm^{-1}}$. A more extensive investigation of this unidentified carrier will be presented by Das et al. (in prep).}

We place a {conservative} upper limit on the C$_2$H$_6$ column density using the local $3\sigma$ uncertainty of the continuum-normalized MRS spectrum {in a region free of absorption features} near the central Q branch of the $\nu_9$ band, where the model predicts the strongest C$_2$H$_6$ absorption. The local $3\sigma$ level is {0.56\%} in normalized intensity. Adopting the same $b$, $f_c$, and 1$\sigma$ $T_{\rm ex}$ range for CH$_3$ ($b=10$~km~s$^{-1}$, $f_c=0.10$, and $T_{\rm ex}$=190$^{+25}_{-20}$~K), we obtain log$[N_{\rm tot}\rm{(C_2 H_6)/\rm{cm}^{-2}}]\lesssim{17.64}^{+0.04}_{-0.05}$, or $N_{\rm tot}({\rm C}_2{\rm H}_6)/N_{\rm tot}({\rm CH}_3)\lesssim{0.8}$.

\section{Discussion} 
  
\subsection{Implications for the Hydrocarbon Chemistry of SMP\,LMC\,011} 
  
The detection of CH$_3$ in SMP\,LMC\,011 has significant implications for our understanding of the formation of small hydrocarbons in C-rich circumstellar environments. The methyl radical occupies a uniquely central position in the reaction network of the PDR associated with the dense torus. CH$_3$ is simultaneously a product of photodissociation, a building block for more complex molecules, and a reactive intermediate that bridges aliphatic and aromatic chemistry. A more complete analysis of the molecular inventory of SMP\,LMC\,011 based on the JWST MIRI/MRS data, including detailed spectral modelling of all detected species, is presented by Das et al.\ (in prep.). Here we focus on the specific chemical implications of the CH$_3$ detection. 
  
\subsection{The origin of the CH$_3$ reservoir: grain erosion in a warm, irradiated torus?} 
  
The detection of CH$_3$ in SMP\,LMC\,011 {provides the first secure, quantitative measurement of gas-phase CH$_3$} in highly evolved C-rich stars. {Whether CH$_3$ is comparably abundant in} the prototypical C-rich AGB star IRC\,+10216 {and} the PPN AFGL\,618 (the two sources with comparable molecular richness) {cannot presently be established. The diagnostic $\sim$16.5 $\mu$m CH$_3$ $\nu_2$ band region is compromised in the archival ISO/SWS spectra of both sources by strong fringing or by its location at the edge of two segments; so these data can neither confirm nor exclude CH$_3$ at column densities comparable to, or below, that derived here.} In IRC\,+10216, the circumstellar chemistry is dominated by C$_2$H$_2$, HCN, and carbon-chain radicals (C$_n$H), while in AFGL\,618, the UV field from the central star drives the photopolymerization of acetylene into polyynes, cyanopolyynes, and methylpolyynes; the available carbon is channelled exclusively through unsaturated, progressively dehydrogenated species \citep{cernicharo2004}. In both {cases}, current photochemical models {of these environments} predict a negligible steady-state {CH$_3$} abundance {(see below)}.
  
The CH$_3$ column density derived here, $N_{\rm tot}(\mathrm{CH_3}) \simeq 5.6\times10^{17}$ cm$^{-2}$, {establishes} CH$_3$ as {an abundant reactive intermediate in the hydrocarbon chemistry within the torus}. Pure gas-phase photodissociation of CH$_4$, a minor but established constituent of C-rich AGB envelopes \citep[e.g.,][]{keady1993}, cannot alone account for the observed CH$_3$ abundance. In the photochemical models of \citet{cernicharo2004}, CH$_3$ is rapidly photodissociated and the {neutral} reactions that could replenish it have low rates at the standard 300\,K reference temperature used in those models, implying a negligible steady-state abundance. {This suggests that an additional source of CH$_3$ may be required in the torus. Independently, SMP~LMC~011 shows C-H absorption associated with aliphatic material (i.e., hydrogenated amorphous carbon; HAC-like), while laboratory experiments demonstrate that energetic processing of such material can release CH$_3$ and other small hydrocarbons (see below)}. We therefore {suggest that erosion of HAC-like} dust grains by UV photons from the central star (or binary system; Paper~\textsc{i}) and/or by shocks driven by the fast ($\sim$122~km\,s$^{-1}$) bipolar outflow~\citep[][]{malek2012} {provides a physically motivated additional source of CH$_3$. The dense, warm torus of SMP\,LMC\,011, directly irradiated by a hot central star \citep[$>$35 kK;][Paper~\textsc{i}]{bernardsalas2006,Shaw2006}, may provide the favourable conditions for efficient grain surface erosion and release of radicals into the gas phase.}

{Our suggested interpretation} is consistent with the detection of aliphatic (HAC-like) C-H stretching modes in absorption at 3.4--3.5 $\mu$m toward SMP LMC 011 together with the lack of any PAH-related emission (Paper~\textsc{i}). Moreover, recent laboratory simulations of dust formation under AGB star conditions demonstrate that HAC-like grains with C-H bonds in aliphatic CH$_2$/CH$_3$ groups are the expected dust product of C-rich AGB envelopes \citep{tajuelocastilla2026}, thus providing the carbonaceous precursor material for UV and/or shock-driven CH$_3$ release in the subsequent PPN phase. {As mentioned above,} laboratory experiments show that UV irradiation and sputtering of HAC grains efficiently release aliphatic CH groups and small radical fragments, primarily CH$_3$ and CH$_4$, into the gas phase \citep[e.g.,][]{Alata2015,duley2015,dartois2017} as the C-C bonds are preferentially photocleaved over the C-H bonds \citep{tajuelocastilla2024}. 

Interestingly, the molecular inventory of SMP\,LMC\,011 shows a striking resemblance to that of the ultra-luminous infrared galaxy IRAS\,07251$-$0248, in which \citet{garciabernete2026} recently detected the same set of hydrocarbons at similar excitation temperatures, attributing their high abundances to erosion of HAC grains or PAHs by cosmic rays. In SMP\,LMC\,011, grain erosion would be instead driven by UV photons and/or shocks. The non-detection of C$_2$H$_6$ in both sources {could suggest} that CH$_3$ is preferentially channelled into reactions with unsaturated radicals rather than predominantly undergoing self-recombination (see below). This may point to similar methyl radical chemistry in grain-erosion environments despite different physical drivers.

\subsection{CH$_3$ as the key intermediate: from grain erosion to benzene and alkyl-substituted aromatics} \label{subsec:discussion_ch3}
  
From the non-detection of C$_2$H$_6$, we derive a {conservative} upper limit of $N_{\rm tot}(\mathrm{C_2H_6})/N_{\rm tot}(\mathrm{CH_3}) \lesssim {0.8}$. {This limit alone does not fix the branching of CH$_3$, since the CH$_3$ ~+~CH$_3$ ~$\rightarrow$~C$_2$H$_6$ stoichiometry caps the C$_2$H$_6$ yield at half the CH$_3$ consumed and any C$_2$H$_6$ formed may itself be photodissociated.} In the dense torus, CH$_3$ produced by grain erosion and CH$_4$ photolysis {may find} two possible fates {(the full network is summarized in Figure~\ref{fig:network})}: self-recombination (CH$_3$ ~+~CH$_3$ ~$\rightarrow$~C$_2$H$_6$) and reaction with the unsaturated radicals (C$_{2n}$H). {However, self-recombination seems to be disfavoured on physical grounds. It has no exothermic bimolecular exit channel \citep[CH$_3$ ~+~CH$_3$ ~$\rightarrow$~C$_2$H$_5$ ~+~ H is strongly endothermic and opens only at temperatures of $\sim$1000-2500~K;][]{stewart1989}, so the C$_2$H$_6$ adduct must be stabilized by a collision or by photon emission. The density of the torus \citep[n$(H_2$) $\approx$ 10$^{7}$ cm$^{-3}$ in the analogous AFGL 618 torus,][]{cernicharo2001b,cernicharo2004} is far below that at which three-body stabilization becomes competitive for such neutral-neutral associations \citep[e.g.,][]{Herbst1980,Herbst2021}, so only the pressure-independent radiative channel remains. Radiative association becomes efficient at low pressure only for sufficiently large adducts \citep[e.g., those with roughly four or more carbon atoms;][]{Vuitton2012} and has a strong inverse temperature dependence \citep[e.g.,][]{Tennis2021}; for an adduct as small as C$_2$H$_6$ at the dust temperature \citep[$\simeq$425 K,][]{malek2012} of the torus, it is therefore expected to be inefficient. Reactions of CH$_3$ with unsaturated radicals, by contrast, access exothermic, largely barrierless bimolecular channels and are not so bottlenecked. Indeed, CH$_3$ additions to radicals such as phenyl are estimated to be much faster than CH$_3$ self-recombination \citep[e.g.,][]{Vuitton2012}. Although reaction-specific rate coefficients for CH$_3$ ~+~C$_{2n}$H are not yet available, the same rapid behaviour is measured experimentally for the analogous unsaturated radical C$_2$H$_3$ \citep[e.g.,][]{fahr1991}. We thus suggest} that reactions with unsaturated radicals {dominate the fate of CH$_3$, placing it} at the root of a rich and branching chemical network {(Figure~\ref{fig:network})}. 
  
The first branch of this network is the methylpolyyne pathway. The reaction of CH$_3$  with C$_{2n}$H radicals {is expected to} produce the methylpolyynes CH$_3$C$_{2n}$H, of which propyne (CH$_3$C$_2$H, $n=1$) is detected in SMP\,LMC\,011 \citep[][and Das et al., in prep.]{malek2012}. Its significantly higher abundance compared to AFGL\,618 \citep{malek2012} {may be} a direct consequence of {a large} CH$_3$ reservoir {in SMP\,LMC\,011}; i.e., at every  step of the polyacetylenic ladder, CH$_3$ competes with C$_2$H$_2$ for the available C$_{2n}$H radicals, redirecting carbon from the pure polyyne chain into the methylpolyyne branch. 
  
The second and chemically more significant branch is the aromatic pathway. SMP\,LMC\,011 harbours an exceptionally large benzene column density, with a C$_6$H$_6$/C$_2$H$_2$ ratio $\sim$20 times higher than in AFGL\,618 \citep{malek2012}. The origin of the anomalously high C$_6$H$_6$/C$_2$H$_2$ ratio remains an open question\footnote{Such a discrepancy remains a puzzle that neither the neutral acetylenic polymerization route of \citet{cernicharo2004} nor the ionic benzene formation route \citep[e.g.,][]{woods2002, woods2003} has been able to fully explain \citep[see][]{malek2012}.}; we note however that benzene destruction via CN could be inhibited in SMP\,LMC\,011 by the marginal HCN abundance \citep[{the parent species of CN, as suggested by}][]{malek2012}, which may contribute to benzene accumulation independent of the formation route. 

Once benzene is formed, the abundant CH$_3$ drives its growth toward alkyl-substituted aromatics through a hydrogen abstraction {(here driven by UV photodissociation rather than thermally) followed by barrierless} methyl addition \citep[see][and references therein]{santoro2020}. For example, 
{VUV photodissociation of benzene yields the phenyl radical (C$_6$H$_5$) as its dominant channel \citep[e.g.,][]{kislov2004}, which then reacts with CH$_3$ in a barrierless step} to form toluene (C$_7$H$_8$); further methyl additions to larger {photochemically generated} aromatic radicals propagate the growth of alkyl-substituted PAHs \citep[see Fig. 11 in][]{santoro2020}. {Similarly, \textit{ortho}-benzyne (o-C$_6$H$_4$) can form via several channels: the barrierless, exoergic C$_2$H\,+\,C$_4$H$_4$ reaction \citep[provided vinylacetylene is available;][]{zhang2011}, H-loss from phenyl, or (to a minor extent) H$_2$-loss from benzene, and then} reacts with CH$_3$ in a barrierless addition step to form the benzyl radical (C$_7$H$_7$), with subsequent hydrogen-atom loss yielding five-membered ring species \citep{bouwman2023}. 

Both CH$_3$-driven pathways simultaneously consume benzene, producing more complex aromatic species and creating a self-amplifying cycle in which the large benzene reservoir is continuously processed toward higher molecular complexity. SMP\,LMC\,011 is therefore a unique target for the detection of the simplest alkyl-substituted aromatics such as toluene (C$_7$H$_8$) and 
ethylbenzene (C$_8$H$_{10}$) in an evolved star. However, their detection with JWST/MIRI is currently hampered by the absence of gas-phase ro-vibrational linelists at astrophysically relevant temperatures. New laboratory spectroscopic measurements or theoretical linelist computations for these species are therefore urgently needed.
  

 \section{Summary and Conclusions} 
  
We report the first {secure} detection of the methyl radical CH$_3$ in a C-rich evolved star SMP\,LMC\,011. CH$_3$ is well fitted by an excitation temperature of $T_{\rm ex} \simeq$~190\,K and a column density of $N_{\rm tot}(\mathrm{CH_3}) \simeq 5.6\times10^{17}$ cm$^{-2}$. We also report the non-detection of ethane (C$_2$H$_6$), {consistent with} CH$_3$ {reacting} preferentially with unsaturated radicals rather than recombining with itself. {The high CH$_3$ abundance may require an additional source, which we tentatively attribute to} the erosion of carbonaceous (HAC-like) dust grains by UV photons and/or shocks {in the dense, warm, directly irradiated torus}; i.e., with grain erosion acting as a direct driver of aromatic chemistry in evolved stars. Once benzene is formed through the acetylenic and/or ionic routes, the abundant CH$_3$ would drive its growth toward alkyl-substituted aromatics (e.g., through hydrogen abstraction and methyl addition) and/or five-membered ring species, with profound implications for PAH formation. SMP\,LMC\,011 is therefore a unique target for searching for alkyl-substituted aromatics, and five-membered ring species in evolved stars, which would provide direct confirmation of CH$_3$-driven aromatic growth, provided that gas-phase ro-vibrational linelists for these species become available. Our results thus call for updated chemical models incorporating {CH$_3$-addition} reactions to understand PAH formation in C-rich circumstellar environments.

\begin{acknowledgments}
J. L., D.A.G.H. and A.M. acknowledge support from the State Research Agency (AEI) of the Spanish Ministry of Science, Innovation, and Universities (MICIU) of the Government of Spain, and the European Regional Development fund (ERDF), under grant PID2023-147325NB-I00/AEI/10.13039/501100011033. This publication is based upon work from COST Action CA21126 - Carbon molecular nanostructures in space (NanoSpace), supported by COST (European Cooperation in Science and Technology). This work is based on observations made with the NASA/ESA/CSA James Webb Space Telescope. All of the data presented in this article were obtained from the Mikulski Archive for Space Telescopes (MAST) at the Space Telescope Science Institute. The data of this specific observing program can be accessed via {\bf the JWST Proto-PAH project (DOI:10.17909/zs5n-2t21).} D.~D., J.~C., E.~P., C.~B., and N.~C. acknowledge support from the Canadian Space Agency (CSA) [24JWGO3B01], and the Natural Sciences and Engineering Research Council of Canada. G.C.S., H.L.D., K.E.K., R.S., N.S., and G.M.W. were supported in part for program 4678 through grants from the STScI under NASA contract NAS5-03127. R.S.'s contribution to the research described here was carried out at the Jet Propulsion Laboratory, California Institute of Technology, under a contract with NASA (80NM0018D0004). 
\end{acknowledgments}





%
\facilities{JWST(MIRI)}




\appendix 

\begin{figure}[h]
    \centering
    \includegraphics[width=\textwidth]{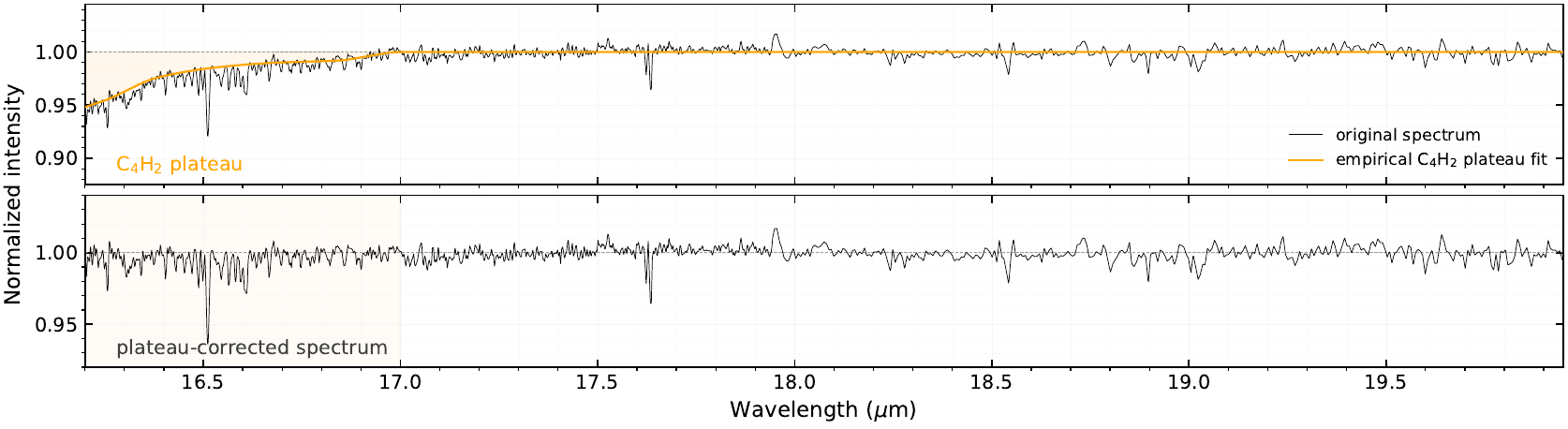}
    \caption{Empirical removal of the broad C$_4$H$_2$ plateau from the 16.2--20~$\mu$m spectrum. \textit{Top panel}: original continuum-normalized JWST/MIRI MRS spectrum of SMP LMC~011, with the smooth empirical baseline adopted for the C$_4$H$_2$ plateau shown in orange. This baseline was obtained by fitting a smooth spline to the binned broad short-wavelength residual after subtracting the representative CH$_3$ model, and is not intended as a physical C$_4$H$_2$ model. \textit{Bottom panel}: plateau-corrected normalized spectrum used for the CH$_3$ model comparison in Figure~\ref{fig:ch3_c6h2_model_v2}.}
    \label{fig:ch3_c6h2_model_plateau_removel}
\end{figure}

Figure~\ref{fig:ch3_c6h2_model_plateau_removel} shows the region covering the C$_4$H$_2$ plateau subtracted in Figure~\ref{fig:ch3_c6h2_model_v2}.

\section{CH$_3$ Fitting Constraints and Parameter Uncertainties}\label{app:uncertainties}

For the spectral fitting of CH$_3$, a unique best-fitting model is difficult to identify using a conventional channel-by-channel $\chi^2$ fitting alone. Strong or broad features dominate the fitting process, but they are affected by saturation, blending, and parameter degeneracies. Their absorption depths no longer increase linearly with column density, while they may still exhibit apparently Gaussian-like profiles at the MIRI resolving power. In contrast, weaker and more optically thin features contribute relatively little to the total statistic~\citep{li2024}, even though they provide more direct constraints on the column density. Robust constraints on the intrinsic line width, $b$, and covering fraction, $f_c$, would ideally come from a simultaneous analysis of multiple molecular species or spectral bands, but such an analysis will be presented in Das et al. (in prep.). We therefore focus on characterizing the degeneracies among the model parameters and identifying the parameter ranges that remain consistent with the observed CH$_3$ spectrum.

To do so, we evaluate the CH$_3$ models using a fixed set of diagnostically important feature windows, each centered on an individual absorption minimum identified in the observed spectrum. Each window spans 1.5 instrumental resolution elements at {the local wavelength-dependent resolving power $R(\lambda)$ adopted in Section~\ref{sec:analysis}}, which is broad enough to sample the local absorption feature over multiple spectral channels while remaining narrow enough to limit contamination from neighboring spectral structure. We then calculate a feature-based statistic for the selected windows. This statistic has the form of a $\chi^2$, but is used primarily as an empirical measure of how well each model reproduces the relative depths of the observed features.

\begin{figure}[h]
    \centering
    \includegraphics[width=0.9\textwidth]{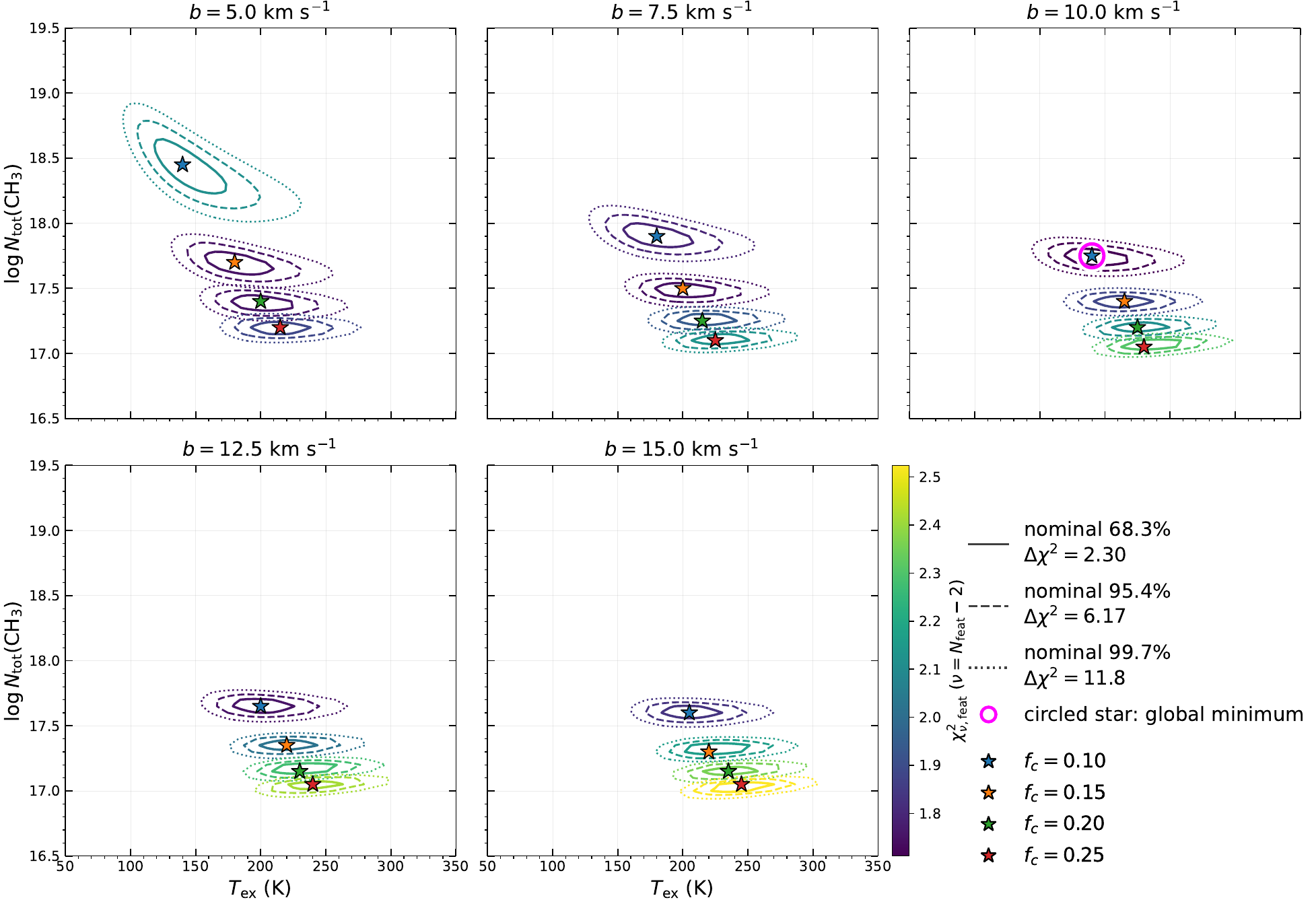}
    \caption{Joint constraints on the CH$_3$ excitation temperature, $T_{\rm ex}$, and column density, $N_{\rm tot}$(CH$_3$), for fixed values of the intrinsic line width, $b$, and covering factor, $f_c$. Each panel corresponds to a different value of $b$, while the colored stars mark the minimum feature-based statistic for the four adopted values of $f_c$. The solid, dashed, and dotted contours represent the nominal joint 68.3\%, 95.4\%, and 99.7\% regions, corresponding to $\Delta\chi^2$=2.3, 6.17, and 11.8 relative to the minimum within each fixed-($b, f_c$) slice. The contour color indicates the reduced feature-based statistic at the slice minimum, with lower values corresponding to better fits. The open circle marks the global minimum over the full grid.}
    \label{fig:NT_contour}
\end{figure}

Figure~\ref{fig:NT_contour} presents the resulting constraints in the $T_{\rm ex}$--log$N_{\rm tot}$ plane for fixed values of $b$ and $f_c$. Although all four parameters could formally be fitted simultaneously, the lack of constraints on $b$ and $f_c$ as well as the continuum uncertainties make the resulting posterior sensitive to the adopted priors and error model. We therefore use a grid-based analysis to display the parameter degeneracies directly. Specifically, our exploration shows that models outside $b$=5-15~km~s$^{-1}$ or $f_c>$0.25 provide poor matches to the observed spectral profiles: larger values of $b$ excessively broaden and blend the CH$_3$ features, smoothing out the observed local minima. In addition, larger values of $f_c$ require column densities that are too low to reproduce the weaker absorption features and their relative depths. We note that the covering fraction also has a direct lower limit of approximately $f_c$=0.075 within the adopted partial-coverage model, as set by the depth of the strongest observed CH$_3$ absorption. Even for a fully saturated line, the maximum absorption depth cannot exceed $f_c$. We therefore examine the joint constraints on $T_{\rm ex}$ and log$N_{\rm tot}$ for fixed ($b$, $f_c$) pair. 

Figure~\ref{fig:NT_contour} reveals several trends in the model parameters: (1) $f_c$ and $N_{\rm tot}$ are strongly degenerate. As $f_c$ increases, a smaller column density is required to reproduce the same absorption depth, as expected from the approximate scaling of optically thin absorption with $f_cN$; (2) the preferred column density does not decrease without limit at large $f_c$, because a sufficiently large column is still required to reproduce the weaker features and the relative depths among features with different optical depths; (3) larger values of $b$ generally produce poorer fits, indicating that excessively broad profiles smooth out the observed feature structure and increase blending among neighboring transitions; and (4) temperature is less tightly constrained than the other parameters. Most of the competitive solutions remain near $T_{\rm ex}\sim200$~K, while modest changes in $T_{\rm ex}$ can be compensated by corresponding adjustments in $N_{\rm tot}$.

Since several combinations of $T_{\rm ex}$, $N_{\rm tot}$, $b$, and $f_c$ produce comparably good fits, the global minimum should not be interpreted as a unique physical solution. We nevertheless adopt it as a representative best-fitting model and quote the corresponding joint $T_{\rm ex}$--log$N_{\rm tot}$ constraints for the fixed ($b$, $f_c$) pair. The minimum feature-based statistic is obtained for $T_{\rm ex}$=190~K, log$[N_{\rm tot}\rm{(CH_3)/\rm{cm}^{-2}}]$=17.75, $b$=10~km~s$^{-1}$, and $f_c=0.1$. For this fixed ($b,f_c$) pair, the nominal joint 68.3\% region gives $T_{\rm ex}=190^{+25}_{-20}$~K and log$[N_{\rm tot}\rm{(CH_3)/\rm{cm}^{-2}}]=17.75^{+0.05}_{-0.05}$. These ranges are the projections of the joint two-dimensional $T_{\rm ex}$--log$N_{\rm tot}$ contour corresponding to $\Delta\chi^2=2.3$. We note that since the selected features are not guaranteed to be statistically independent and their uncertainties include an empirical continuum removal, we tested the nominal $\Delta\chi^2=2.3$ threshold with slice-by-slice parametric Monte Carlo simulations and verified that the calibrated 68.3\% thresholds were close to the nominal two-parameter value. {These confidence regions should therefore not be interpreted as the full uncertainty on the physical parameters, but rather as conditional constraints within each adopted ($b$, $f_c$) model.}

\section{CH$_3$ chemical network in the warm, UV-irradiated torus of SMP\,LMC\,011}

\begin{figure}[h]
    \centering
    \includegraphics[width=1.1\textwidth]{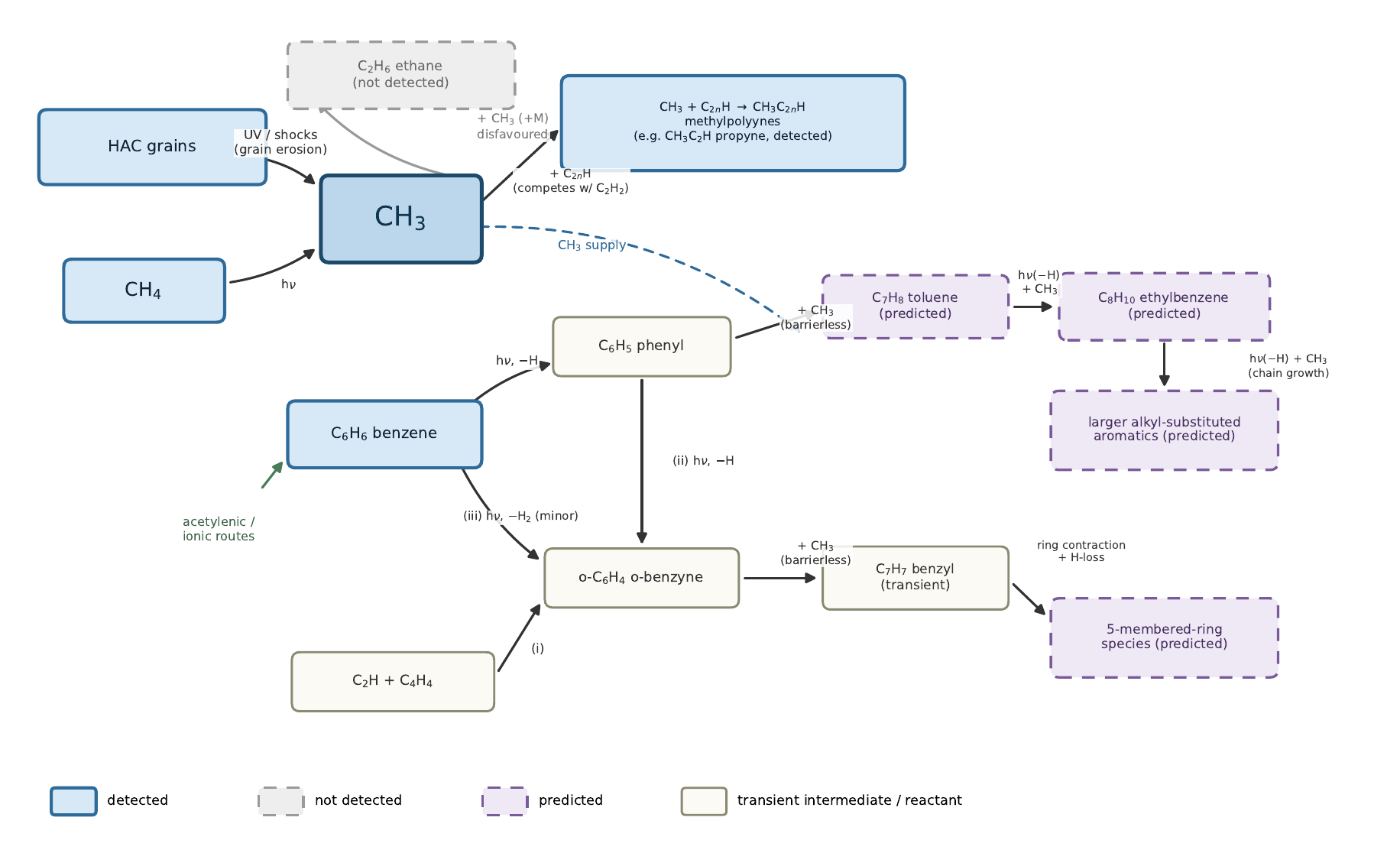}
    \caption{Simplified schematic of the CH$_3$ chemical network in the warm, UV-irradiated torus of SMP\,LMC\,011 (see text).}
    \label{fig:network}
\end{figure}

{Figure~\ref{fig:network} displays a simplified schematic of the CH$_3$ chemical network in the warm, UV-irradiated torus of SMP\,LMC\,011. Species are colour-coded: blue = detected in this source (or, for the methylpolyynes, detected via CH$_3$C$_2$H); grey dashed = searched for but not detected; purple dashed = predicted but not yet detected; cream = transient reactive intermediates or reactants. CH$_3$ is supplied by erosion of HAC grains (UV photons and/or shocks) and, more minorly, by CH$_4$ photodissociation. It then follows three competing routes: (i) self-recombination to C$_2$H$_6$ (disfavoured at the low density of the torus; see Subsection~\ref{subsec:discussion_ch3}); (ii) addition to unsaturated carbon-chain radicals C$_{2n}$H to form methylpolyynes (CH$_3$C$_{2n}$H; e.g.\ propyne, CH$_3$C$_2$H), in competition with C$_2$H$_2$; and (iii) an aromatic-growth branch. In the latter, VUV photodissociation of benzene generates the phenyl radical (C$_6$H$_5$) by H-atom loss, to which CH$_3$ adds barrierlessly to form toluene (C$_7$H$_8$);
successive H-abstraction/photolysis and CH$_3$ additions then build ethylbenzene
(C$_8$H$_{10}$) and larger alkyl-substituted aromatics. In parallel, \textit{ortho}-benzyne (\textit{o}-C$_6$H$_4$) may form via (i) the barrierless
C$_2$H\,+\,C$_4$H$_4$ reaction, (ii) H-loss from phenyl, or (iii) direct H$_2$-loss from benzene (a minor channel). The barrierless reaction \textit{o}-C$_6$H$_4$\,+\,CH$_3$ then yields the transient benzyl radical (C$_7$H$_7$), which by ring contraction and H-atom loss produces five-membered-ring species \citep[e.g., fulvenallene and ethynylcyclopentadienes;][]{bouwman2023}. Note that benzene itself is supplied by acetylenic/ionic routes rather than by CH$_3$ (see Subsection~\ref{subsec:discussion_ch3}).}

\bibliography{biblio_paper_ApJL_LMC11_CH3}{}
\bibliographystyle{aasjournal}



\end{document}